\documentclass[draft,showkeys,showpacs,eqsecnum,nofootinbib,aps]{revtex4}
\def\be{\begin{equation}}
\def\ee{\end{equation}}
\def\bea{\begin{eqnarray}}
\def\eea{\end{eqnarray}}

\def\be{\beta}

\def\tr{{\mbox{\,tr}}}

\def\tr{{\mbox{\,tr}}}

\def\I_M{{I_{\scriptscriptstyle M\times M}}}

\begin{document}

\title{\vskip -60pt
{\small\begin{flushright}  {\tt hep-th/0610067}\phantom{aaaaaaaaa}
\end{flushright}}
\vskip 45pt Supersymmetric Q-Lumps in the Grassmannian nonlinear
sigma models } \vspace{4.0cm}
\author{Dongsu Bak${}^{a}{}%^{\natural}
$, Sang-Ok Hahn${}^{b}$,
Joohan Lee${}^{a}{}%^{\dagger}
$, and Phillial Oh${}^{b}
%{}^\ddagger
$} \affiliation{${}^{a}$ Department of Physics,
University of Seoul, Seoul 130-743 Korea}
\affiliation{${}^{b}$Department of Physics and Institute of Basic
Science,  Sungkyunkwan University, Suwon 440-746 Korea}
\date{\today}%\\
%\email{{\dagger}joohan@kerr.uos.ac.kr, **ploh@dirac.skku.ac.kr}
\vspace{3.0cm}
\begin{abstract}
We construct the  ${\cal N}=2$ supersymmetric Grassmannian nonlinear
sigma model for the massless case and extend it to massive  ${\cal
N}=2$ model
%of the supersymmetries
by adding an appropriate superpotential. We then
study their BPS equations leading to supersymmetric
Q-lumps carrying both %the
topological and %the
Noether
charges. These solutions are shown to be always time
dependent even sometimes involving multiple
frequencies. Thus we illustrate explicitly  that the
time dependence is consistent with remaining supersymmetries
of solitons.

%We study time-dependent BPS solitons in supersymmetric
%Grassmannian nonlinear $\sigma$-model. In the presence of a
%potential term, the model admits Q-lump solutions. Supersymmetric
%properties of the solutions are discussed.
\end{abstract}
~\newline ~\newline ~\newline
%%
%%\pacs{04.60.K, 04.70.Dy, 11.25.Hf}
%%
%%\narrowtext
%%%
%~\newline $\overline{\mbox{ E-mail

 \maketitle

\vspace{8.0cm}

$\overline{\mbox{
E-mail
addresses\,\,:}~%{}^{\natural}
\mbox{dsbak@mach.uos.ac.kr},\ %{}^{\dagger}
\mbox{joohan@kerr.uos.ac.kr},\  %{}^{\ddagger}
\mbox{ploh@dirac.skku.ac.kr}~~~}$
\thispagestyle{empty}
\newpage
\setcounter{footnote}{0}

\baselineskip 20pt

%%%%%%%%%%%%%%%%%%%%%%%%%%%%%%%%%%%%%%%%%%%%%%%%%%%%%%%%%%%%%%%%%%%
\section{Introduction}
%%%%%%%%%%%%%%%%%%%%%%%%%%%%%5%%%%%%%%%%%%%%%%%%%%%%%%%%%%%%%%%%%%%

$Q$-lumps \cite{lees} are topological soliton solutions which also
carry a conserved Noether charge $Q$ \cite{cole} in a class of
nonlinear sigma models of massive K\"{a}hler or hyper-K\"{a}hler
models \cite{abra}. Unlike the pure topological solitons which are
unstable against the size perturbation \cite{derr}, these
configurations are prevented from collapsing through time-dependent
internal rotations and the size is determined by the conserved
Noether charge. It is essential to have a  potential term of specific
form which is just a mass term in the linearized theory
% to be included in the nonlinear sigma model Lagrangian 
 and it is known
that, for a K\"{a}hler sigma model, $Q$-lumps can 
exist only if the target
manifold has a continuous symmetry with at least one fixed point
\cite{abra}.

For any given value of $Q$, this  potential term enables
the existence the
BPS %Bogomolny
bound, which
%and it
guarantees that the $Q$-lumps minimize their energy.
% for any
%given value of $Q$.
It is known that the specific potential
naturally arises from the supersymmetric generalization of the
bosonic nonlinear sigma model \cite{abra}.

%The fact that the
%solution is stationary allows to  preserves fraction of the
%supersymmetry
Q-lumps %These solutions
involve a nonvanishing kinetic contribution due to their
time dependence
%and these
and such configurations  attracted a great deal of
recent interest in the study of  %time-dependent
BPS solutions with nontrivial kinetic terms 
in both field
theory and string theory \cite{gaun,eto,townsend}.
On the other hand, more {\it
explicit} construction of the supersymmetric generalization of
the original massive $O(3)$ \cite{lees} and the K\"{a}hler
\cite{abra} nonlinear sigma model solitons in 2+1 dimensions seems to be
lacking. In view of the growing interest on the subject, it would
be desirable to investigate the roles  of the supersymmetries
in detail.
%in this direction.

In this note, we study the supersymmetric Grassmannian nonlinear
sigma model  in 2+1 dimension \cite{pere} and its $Q$-lump
solutions. We first construct ${\cal N}=2$ supersymmetric massless
model from ${\cal N}=1$ superfield formalism  in constrained variable
approach by eliminating the auxiliary fields. We then extend it to
the massive ${\cal N}=2$ model
%with introduction of
%the ${\cal N}=1$
by adding an appropriate
superpotential term\footnote{There have been some constructions of 
d=2 ${\cal N}=2$ massive nonlinear sigma models \cite{Gates}.  
See also Ref.~\cite{Nitta} for the d=4 ${\cal N}=2$ massive 
sigma models.
}.  
The corresponding sets of BPS equations can be studied either
by the method of  completing squares or by directly
finding conditions for the remaining supersymmetries.
The former leads to 
sets of conditions for the saturation of the BPS
energy bound by charges while the latter
leading to conditions
%dictates %
%follow from the
%set of
%by requiring that a bosonic configuration 
for preserving a fraction of
supersymmetries. We shall find  that the sets of BPS equations from
the former are more general than those from the latter unlike the
cases of previously known examples. The resulting  Q-lump solutions are
always time dependent. Thus this illustrates the consistency of 
time  dependence of solutions and 
 remaining supersymmetries. We also discuss
the supersymmetric multiply charged Q-lump solution that involves
time dependence of many frequencies. The existence of such solutions
is highly nontrivial since the  sectors of different frequencies are
interacting with each other.

The kinetic energy due to the time dependence cannot be relaxed at
least classically due to the BPS bound and the conservation of the
topological and electric charges. When there are enough number of
remaining supersymmetries, the kinetic energy can be protected even 
from the quantum corrections.

In Section II, we will set up our notations and introduce the ${\cal
N}=2$ massless Grassmannian nonlinear sigma model. In Section III,
we extend the massless model to the massive one with ${\cal N}=2$
supersymmetries by adding an appropriate superpotential. We then
study the BPS equations by completing squares. In Section IV, we
obtain conditions for the remaining supersymmetries leading to 1/2
BPS states and discuss solutions of BPS equations.  In Section V,
the multi charged Q-lump solutions  involving  many frequencies  are
constructed. Last section is devoted to concluding  remarks.

%The BPS
%equations are given and it is shown that the $Q$-lumps are 1/2 BPS
%states. We also find that $Q$-lumps allow configuration where
%they consist of many different $Q$-charges.

%%%%%%%%%%%%%%%%%%%%%%%%%%%%%%%%%%%%%%%%%%%%%%%%%%%%%%%%%%%%%%%%%%%
\section{Supersymmetric Grassmannian
Model in Three Dimensions}
%%%%%%%%%%%%%%%%%%%%%%%%%%%%%5%%%%%%%%%%%%%%%%%%%%%%%%%%%%%%%%%%%%%
In this section we introduce the ${\cal N}=2$ supersymmetric
Grassmannian nonlinear sigma model for the massless case.  It is
basically a
 nonlinear sigma model with a target
space of the Grassmannian manifold but also possesses the ${\cal
N}=2$ extended supersymmetries. To make this note self contained,
let us begin by  setting up some notations.
%This section is included mainly for setting the notation and to
%make the paper self-contained.
 The superspace $Z=(x,\theta)$ is
given by a spacetime coordinates $x$ and an anticommuting
coordinates $\theta$ which is a two-component Majorana spinor
$\theta$ \cite{gate}.

In   2+1 dimensions, the Dirac algebra is given by three $2\times
2$ matrices ${\gamma}^{\mu}$ with
\begin{eqnarray}
{\gamma}^{0} = {\sigma}_{2}\,, \,\,\,\,\, {\gamma}^{1} =
i{\sigma}_{3}\,, \,\,\,\,\, {\gamma}^{2} = i{\sigma}_{1}\,,
\end{eqnarray}
where the $\sigma$'s denote the Pauli matrices and the index
$\mu$ runs over $ 0, 1, 2$.
They satisfy the
%commutation relations
Clifford algebra,
\begin{eqnarray}
\lbrace \gamma^{\mu}  , \, \gamma^{\nu}\rbrace =
2\eta^{\mu\nu}\,,
%\,\,\,\,\, \lbrack \gamma^{\mu} \,\, , \,\,
%\gamma^{\nu}\rbrack = -2i\epsilon^{\mu\nu\rho} \gamma_{\rho} \,,
\end{eqnarray}
where  ${\eta}^{\mu\nu}$ is the  Minkowski metric of signature
$(1 \,, \,\,\, -1\,,
\,\,\, -1)$. We introduce the two-component Majorana spinor
$\theta$
\begin{eqnarray}
\theta = \left[
\begin{array}{c}
\theta_1 \\
\theta_2
\end{array}
\right] \,, \,\,\,\,\, \bar{\theta} = \left[
\begin{array}{r}
\bar{\theta}^{1}\;\; \bar{\theta}^{2}
\end{array}
\right] \,,
\end{eqnarray}
satisfying the anticommuting relation, $\lbrace {\theta}_{\alpha}
 , \, {\theta}_{\beta} \rbrace = 0 $, where the adjoint of a
spinor is, as usual, $\bar{\theta}=\theta^{\dagger}\gamma^{0}$.
For bosonic variables, we use the notation ${\bar b} =b^\dagger$.

The Grassmannian manifold, $Gr(N,M)$, is the homogeneous space
defined by $U(N+M)/U(N)\times U(M)$. The ${\cal{N}}=1$ scalar
superfield $\Phi$ for the Grassmannian
model %$Gr(N,M)=U(N+M)/U(N)\times U(M)$
can be  written in
component form as
\begin{equation}
\Phi%_{(N\times M)}
(x,\theta) = \phi%_{(N\times M)}
(x) +
\bar{\theta}\psi%_{(N\times M)}
(x) + \frac{1}{2} \bar{\theta}\theta
F%_{(N\times M)}
(x) \,,
\end{equation}
%where it is an $N\times M$ complex matrix with $\phi_{N\times M}$
%being  a complex scalar field, $\psi_{N\times M}$ a Dirac spinor,
%and $F_{N\times M}$ an auxiliary field. 
where every component is an $(N+M)\times M$ matrix valued field.
The action for the
supersymmetric Grassmannian $\sigma$ model is given by
supersymmetrizing the bosonic model \cite{pere};
\begin{eqnarray}
S_0=\int d^3xd^2\theta \frac{1}{2}\tr \bigg\{\overline{\nabla\Phi}
\nabla\Phi + 2\Sigma\left({\bar{\Phi}}\Phi - I_{(M\times
M)}\right) \bigg\}\,,\label{action1}
\end{eqnarray}
where $\nabla_{\alpha}\Phi = D_{\alpha}\Phi - i\Phi A_{\alpha}$ is the
gauge covariant derivative with $D_{\alpha}=\frac
{\partial}{\partial{\bar{\theta}}^{\alpha}}-i(\gamma^{\mu}\theta)_{\alpha}
\partial_{\mu}$ being the supercovariant derivative, and 
$A_{\alpha}$ is a real
$M\times M$ matrix  spinor gauge superfield given in the Wess-Zumino
gauge by
\begin{equation}
A_\alpha=i(\gamma^\mu\theta)_\alpha A_\mu
+\frac{1}{2}\bar\theta\theta\omega_\alpha\,. \end{equation}
The superfield $\Sigma$
denotes an $M\times M$ matrix valued Lagrange multiplier,
\begin{eqnarray}
\Sigma_{(M\times M)} = \sigma + \bar{\theta}\xi +
\frac{1}{2}\bar{\theta}\theta \alpha \,,
\end{eqnarray}
 and the corresponding supersymmetric
constraint is $\bar{\Phi}\Phi = I_{M\times M}$ where
$\I_M$ is the $M\times M$ identity matrix.
In
component forms, this becomes $\bar\phi\phi=1,~
\bar\psi\phi=\bar\phi\psi=0,~\bar\phi F+\bar
F\phi-\bar\psi\psi=0$, which we assume to hold throughout this
note.

The above action (\ref{action1}) is invariant under global
$SU(N+M)$ transformation $\Phi\rightarrow  G\Phi\,, ~ G\in
SU(N+M)$ as well as the  gauge transformation  given with $U\in
U(M)$ as
\begin{equation}
\begin{array}{ll}
\Phi\rightarrow\phi\,{\bar U}\,,~~~&~~~A_{\alpha}\rightarrow
UA_{\alpha}{\bar U}+iU\partial_{\alpha}{\bar U}\,.
\end{array}
\label{gauge}
\end{equation}
The $U(M)$ covariant derivative is defined as
 $D_\mu\phi=\partial_\mu\phi+i\phi A_\mu$ with
the  field strength given by
$F_{\mu\nu}=\partial_{\mu}A_{\nu}-
\partial_{\nu}A_{\mu}-i[A_\mu,\,A_\nu]$.

In component form, the supersymmetric Grassmannian model yields
\begin{eqnarray}
S_0&=&\int d^3x %\frac{N}{g}
~\tr~\Big\{  (D_\mu \bar\phi) D^\mu \phi +i\bar{\psi}\gamma^{\mu}
D_{\mu} {\psi} + \bar{F}F +\frac{i}{2}\bar{\psi}\phi\omega
-\frac{i}{2} \bar\omega\bar\phi \psi\nonumber \\
&&~~~~~~~~~~~~+ \sigma(\bar{\phi}F - \bar{\psi}\psi + \bar{F}
\phi) - \bar{\phi}\bar{\xi}\psi - \bar{\psi}\xi\phi +
\alpha(\bar{\phi}\phi - 1)\Big\}\,. \label{action2}
\end{eqnarray}
By the construction,
the system is invariant under the supersymmetry transformations
\begin{eqnarray}
&&\delta %_{\epsilon}
\phi = \bar{\epsilon}_M \psi \,,
~~~~~~~~~~~~~~~~~~~~~~~~~
\delta %_{\epsilon}
{\bar\phi} = \bar{\psi}\epsilon_M \,, \\
&&\delta %_{\epsilon}
\psi = \epsilon_M F - i\gamma^{\mu}\epsilon_M (
D_{\mu}\phi) \,, ~~~~~~\, \delta %_{\epsilon}
\bar{\psi} = {\bar
F}\bar{\epsilon}_M + i(D_{\mu}
\bar\phi)\bar{\epsilon}_M \gamma^{\mu} \,, \\
&& \delta %_{\epsilon}
F = -i\bar{\epsilon}_M \gamma^{\mu}D_{\mu}\psi +
\frac{i}{2}\bar{\epsilon}_M \phi\omega \,, ~~ \delta %_{\epsilon}
\bar
F = i({D_{\mu}}{\bar\psi})\gamma^{\mu}\epsilon_M
- \frac{i}{2}\bar{\omega}{\bar\phi}\epsilon_M \,, \\
&&\delta %_{\epsilon}
A_{\mu} =
\frac{i}{4}(\bar{\epsilon}_M \gamma_{\mu}\omega
- \bar{\omega}\gamma_{\mu}\epsilon_M) \,, \\
&&\delta %_{\epsilon}
\omega = \frac{1}{2}[\gamma^{\mu} , \,
\gamma^{\nu}] \epsilon_M F_{\mu\nu} \,,
~~~~~~~~\delta %_{\epsilon}
\bar{\omega} =
-\frac{1}{2}F_{\mu\nu}\bar{\epsilon}_M [\gamma^{\mu} , \,
\gamma^{\nu}].\label{susycon}
\end{eqnarray}
where $\epsilon_M$ denotes an anticommuting Majorana spinor.

%In order to obtain the pure BPS equations we consider the
%expression of the energy.
%\begin{eqnarray}
%E&=&\int d^2xT_{00},\\
%&=&\int d^2x
%\tr\left\{(D_0\phi)^\dagger(D_0\phi)+(D_i\phi)^\dagger(D_i\phi)
%\right\},\\
%&=&\left\{\begin{array}{l} \left[\int d^2x
%\tr\left((1-\phi\\bar\phi)AA^\dagger+|A_0-i\\bar\phi\partial_0\phi|^2
%+2 |D_z\phi|^2\right)\right]+
%2\pi T\\
%\left[\int d^2x
%\tr\left((1-\phi\\bar\phi)AA^\dagger+|A_0-i\\bar\phi\partial_0\phi|^2
%+2 |D_{\bar z}\phi|^2\right)\right] -2\pi T\end{array}\right.,
%\end{eqnarray}

One may eliminate
the auxiliary fields %can be eliminated by
using their equations of
motion,
\begin{eqnarray}
F=-\sigma \phi,~ \sigma=-\frac{1}{2}\bar\psi \psi,~
A_\mu=i\bar\phi\partial_\mu\phi+\frac{1}{2}\bar\psi\gamma_\mu\psi\,.
\end{eqnarray}
%Substitution into Eq. (\ref{action2})
%leads to
The resulting action takes the form,
\begin{eqnarray}
S_0 &=&\int d^3 x %\frac{N}{g}
\tr\left\{\vert D_\mu\phi\vert^2 +i\bar\psi\gamma^\mu D_\mu\psi
+\frac{1}{4}(\bar\psi\psi)^2\right\}\nonumber
\\&=&\int d^3 x %\frac{N}{g}
\tr\left\{\vert\partial_\mu\phi\vert^2 +i\bar\psi\gamma^\mu
\partial_\mu\psi
-(i\bar\phi\partial_\mu\phi+\frac{1}{2}\bar\psi\gamma_\mu\psi)^2
+\frac{1}{4}(\bar\psi\psi)^2\right\}.\label{Aoyama-npb}
\end{eqnarray}
Here comes the observation of Refs.~\cite{Aoyama-npb, witten}, which
deal with the case of $CP(n)$ only. Namely the system in
(\ref{Aoyama-npb}) has in fact ${\cal N}=2$ extended
supersymmetries. The transformation rules are almost same as before
but the Majorana spinor $\epsilon_M$ is now replaced by a complex
Dirac spinor $\epsilon$. Explicitly the
  %The above  Eq. (\ref{Aoyama-npb}) has in fact ${\cal N}=2$
%supersymmetry. The case of supersymmetric $CP(N)$ was discussed
%before \cite{Aoyama-npb}. 
${\cal N}=2$ SUSY transformation rules are
\begin{eqnarray}
\delta\phi&=&\bar\epsilon\psi,~\delta\bar\phi=\bar\psi\epsilon,\nonumber\\
\delta\psi&=&\frac{1}{2}\epsilon\phi\bar\psi
\psi-i\gamma^\mu\epsilon\partial_\mu\phi
+\gamma^\mu\epsilon\phi\left(i\bar\phi\partial_\mu\phi+\frac{1}{2}
\bar\psi\gamma^\mu\psi\right),
\label{susycontt}\\
\delta\bar\psi&=&\frac{1}{2}\bar\psi\psi\bar\phi\bar\epsilon+i
\partial_\mu\bar\phi
\bar\epsilon\gamma^\mu
+\left(i\bar\phi\partial_\mu\phi+\frac{1}{2}\bar\psi\gamma^\mu\psi\right)
\bar\phi\bar\epsilon\gamma^\mu. \nonumber
\end{eqnarray}
%Here, $\epsilon$ is a complex Dirac spinor.
Note that, in this description,
%there is
%Since we are working on the system with constarints, the
%supersymmetry transformation should be consistent with the constarint.
%Indeed one may check that, for instance,
%One can check that
the constraint,
$\bar{\phi}\psi=\bar{\psi}\phi=0$, is only solved implicitly
but one can check that this condition
is invariant under the
transformations in
(\ref{susycontt}).

For the study the BPS equations, it is convenient to
introduce the complex
coordinates,
\begin{equation}
z=\frac{1}{\sqrt{2}}(x_{1}+ix_{2})\,,~~~~~\bar{z}=\frac{1}
{\sqrt{2}}(x_{1}-ix_{2})\,,
\end{equation}
and
$\partial_\pm=\frac{1}{\sqrt{2}}(\partial_{1}\mp i\partial_{2})$,
%\bar{\partial}=\frac{1}{\sqrt{2}}(\partial_{1}+i\partial_{2})
and
$A_\pm=\frac{1}{\sqrt{2}}(A_{1}\mp iA_{2})$
%,
%A_{\bar{z}}=\frac{1}{\sqrt{2}}(A_{1}+iA_{2})$.
where $(+, -)$ refer to the holomorphic/antiholomorphic component
of vectors in two spatial dimensions.

In considering BPS configurations, we shall consider purely bosonic
configuration only by setting all the fermionic part to zero.
By completing squares, %The expression of
the Hamiltonian $H$ of the bosonic part only  can be rearranged by
\begin{equation}
\begin{array}{ll}
H &\!=\displaystyle{\int}{\rm
d}^{2}x\,\tr\left(|D_{0}\phi|^2  %)^{\dagger}D_{0}\phi
+|D_+\phi|^2
%)^{\dagger}D_{z}
%\phi
+|D_-\phi|^2 % )^{\dagger}D_{\bar{z}}\phi
\right)
\\
%%%
%%=\displaystyle{\int}{\rm d}^{2}x\,\tr\left[(D_{0}\phi)^{\dagger}D_{0}\phi
%%+(D_{i}\phi)^{\dagger}D_{i}\phi\right]
%%%
%{}&{}\\
%{}&\!=%\left\{\begin{array}{l} \displaystyle{\int}{\rm
%d}^{2}x\,\tr\left[(D_{0}\phi)^{\dagger}D_{0}\phi+2(D_{\bar{z}}\phi)^{\dagger}
%%D_{\bar{z}}\phi\right]-2\pi T\\
%{}\\
%\displaystyle{\int}{\rm
%d}^{2}x\,\tr\left[(D_{0}\phi)^{\dagger}D_{0}
%\phi+2(D_{z}\phi)^{\dagger}D_{z}\ph% i\right]+2\pi
%T
%\end{array}\right.\\
{}&\! =
\displaystyle{\int}{\rm
d}^{2}x\,\tr\left(|D_{0}\phi|^2 +2|D_\mp\phi|^2 \right)
\mp 2\pi T %\\
%{}&{}\\
%{}&\!
\ \geq\  2\pi |T|\,,
\end{array}
\end{equation}
where the topological vortex number $T$, which can be
%can be
written as a
boundary term, is defined by
\begin{equation}
%\begin{array}{ll}
T \equiv \frac{i}{2\pi}
\displaystyle{\int}{\rm
d}^{2}x\,\tr\left(\epsilon^{ij}(D_{i}\phi)^{\dagger}D_{j}\phi\right)
%%=\displaystyle{\int}{\rm
%%d}^{2}x\,\tr\left[i\epsilon^{ij}D_{i}(\phi^{\dagger}D_{j}\phi)
%+F_{12}\right]
%\\
%{}&{}\\
%{}&
=\frac{i}{2\pi}\displaystyle{\oint_{\infty}}{\rm d} x_i %\cdot
{\tr
\left(\phi^{\dagger}\partial_i\phi\right)}\,.
%\end{array}
\end{equation}
%%%
%%With  the ansatz to use shortly, $Q$ will be identified as  the
%%topological quantity without discarding the boundary term.\newline
%%%
Thus for a given sector of the vortex number, the Hamiltonian is
bounded from below. The saturation of the bound occurs if the
BPS/anti-BPS equations,
%The saturation of the energy occurs  when the time-independent BPS
%or anti-BPS equations are satisfied
\begin{equation}
%\begin{array}{lll}
\partial_{0}\phi =0\,, \ \ \ D_\mp \,\phi =0
%{\bar{z}}\phi&\mbox{:\,BPS~equations}\,,\\
%{}&{}&{}\\
%0=\partial_{0}\phi\,,~&0=D_{z}\phi&\mbox{:\,anti-BPS~equations}\,.
%\end{array}
\label{BPSeq}
\end{equation}
are satisfied where we refer the upper/lower signature for the
BPS/anti-BPS sector respectively.

The solutions of these equations correspond to the well known
holomorphic/antiholomorphic vortices of the two dimensional
nonlinear sigma model \cite{pere}, which are obviously time
independent as required by the  BPS equations. We shall not
discuss their properties any further here.

%%%%%%%%%%%%%%%%%%%%%%%%%%%%%%%%%%%%%%%%%%%%%%%%%%%%%%%%%%%%%%%%%%%%%%
%%%%%%%%%%%%%%%%%%%%%%%%%%%%%%%%%%%%%%%%%%%%%%%%%%%%%%%%%%%%%%%%%%%%%%%
\section{Massive ${\cal N}=2$ Model and  Time-dependent BPS Solitons}
%%%%%%%%%%%%%%%%%%%%%%%%%%%%%%%%%%%%%%%%%%%%%%%%%%%%%%%%%%%%%%%%%%%%%%%
%%%%%%%%%%%%%%%%%%%%%%%%%%%%%%%%%%%%%%%%%%%%%%%%%%%%%%%%%%%%%%%%%%%%%%%
In this section, we would like to introduce first the massive Grassmannian
sigma model with ${\cal N}=2$ supersymmetry. One can make the
sigma model massive by the introduction of appropriate
superpotential as we shall explain below. We then find the
corresponding BPS equations 
by the methods of completing
squares of Hamiltonian as in the previous section.

We begin by
 introducing a superpotential %with some positive constant $\alpha$
of the following form,
\begin{eqnarray}
W(\Phi)=\frac{1}{2} %\int  d^2 \theta
\tr~ \omega\, \bar\Phi
P\Phi.\label{potential}
\end{eqnarray}
where $P$ is the $(N+M) \times (N+M)$ Hermitian projection matrix satisfying
$P^2 =P$ and $\omega$ is a real positive number.
Performing the $\theta$ integration, the action in a component form
reads
\begin{eqnarray}
S_1= \int d^3x ~\omega \tr~\left(\bar\phi P F+\bar FP\phi-\bar\psi
P\psi\right).
\end{eqnarray}
 Then let us consider the total action $S=S_0+S_1 $ with $S_0$
 in (\ref{action2}). It has manifest ${\cal N}=1$
 supersymmetry by construction. Eliminating the auxiliary field by using the
 equations of motion,  % as before, we obtain
\begin{eqnarray}
F=\omega P\phi-\phi\sigma,~ \sigma=\omega \bar\phi
P\phi-\frac{1}{2}\bar\psi\psi\,,
A_\mu=i\bar\phi\partial_\mu\phi+\frac{1}{2}\bar\psi\gamma_\mu\psi\,,
\end{eqnarray}
the action becomes
%Substitution into $S_t$ yields
\begin{eqnarray}
S &=&\int d^3 x %\frac{N}{g}
\tr\left\{\vert D_\mu\phi\vert^2 +i\bar\psi\gamma^\mu D_\mu\psi
+\left(\omega \bar\phi
P\phi-\frac{1}{2}\bar\psi\psi\right)^2-\omega^2
\bar\phi P \phi+\omega \bar\psi P\psi\right\}\nonumber\\
&=&\int d^3 x %\frac{N}{g}
\tr\left\{ \vert\partial_\mu\phi\vert^2 +i\bar\psi\gamma^\mu
\partial_\mu\psi
-\left(i\bar\phi\partial_\mu\phi+\frac{1}{2}\bar\psi\gamma_\mu\psi\right)^2
+\left(\omega \bar\phi
P\phi-\frac{1}{2}\bar\psi\psi\right)^2-\omega^2 \bar\phi P
\phi+\omega \bar\psi P\psi\right\}.
\end{eqnarray}
One may then show that
%It can be shown that
the above action is invariant under the
following  ${\cal N}=2$ supersymmetry transformations
\begin{eqnarray}
\delta\phi&=&\bar\epsilon\psi,~~~
\delta\bar\phi=\bar\psi\epsilon~\nonumber\\
\delta\psi&=&\epsilon\left(\frac{1}{2}\phi\bar\psi\psi+\omega
(P\phi-\phi\bar\phi P
\phi)\right)-i\gamma^\mu\epsilon\partial_\mu\phi
+\gamma^\mu\epsilon\phi\left(i\bar\phi\partial_\mu\phi+\frac{1}{2}
\bar\psi\gamma^\mu\psi\right),\\
\delta\bar\psi&=&\left(\frac{1}{2} \bar\psi\psi\bar\phi
 + \omega (\bar\phi P - \bar\phi P
\phi\bar\phi)\right)\bar\epsilon+i\partial_\mu\bar\phi
\bar\epsilon\gamma^\mu +\left(i\bar\phi\partial_\mu\phi+
\frac{1}{2}\bar\psi\gamma^\mu\psi\right)\bar\phi\bar\epsilon\gamma^\mu.
~ \label{susycontt2}
\end{eqnarray}
by a straightforward computation.
In addition,
 one may check that the constraint,
 $\bar{\phi}\psi=\bar{\psi}\phi=0$, is also invariant
under the ${\cal N}=2$
transformations, which insures the consistency of our approach.
% given in Eq. (\ref{susycontt2}).

Note that the theory
% Since the theory
possesses also a global
%$U(N)$ 
symmetry defined by
\begin{eqnarray}
\delta %_{\tilde\epsilon}
\phi=i  P\phi,
~\delta %_{\tilde\epsilon}
\bar\phi=-i\bar\phi P \,.
\end{eqnarray}
%where $u$ denotes $N\times N$ Hermitian constant matrix satisfying
%$[u, P]=0$. 
The expression for  the corresponding conserved current
density is given by
\begin{eqnarray}
J^\mu=i{\rm tr}\Bigl((D^\mu\phi)\bar\phi
P-P\phi(D^\mu\bar\phi)\Bigr)\,,\label{current-a}
\end{eqnarray}
%and the Grassmannian constraint coming from the variation of the
%Lagrangian multiplier
%\begin{eqnarray}
%\bar\phi\phi=1_{M\times M}
%\end{eqnarray}
and 
below we shall make use of the resulting %U(1) component of its 
Noether charge
%defined by %So we obtain the Noether charge,
\begin{eqnarray}
Q=-i\int d^2x \tr \left((D_0\phi)\bar\phi
P-P\phi(D_0\bar\phi)\right)\,.
\end{eqnarray}
Note that the Hamiltonian, including the contribution from the
superpotential,  takes a form,
\begin{eqnarray}
H =\int d^2x
\tr\left\{|D_0\phi|^2 %^\dagger(D_0\phi)
+|D_i\phi|^2 %^\dagger(D_i\phi)
+\omega^2\left(\bar\phi P\phi-(\bar\phi P\phi)^2\right)
\right\}\,,
%\label{inequality}
\end{eqnarray}
where only bosonic part are turned on.

Again by the methods of completing squares,
one part of  Hamiltonian, $H_1$,
%The energy is written by
can be rearranged by
\begin{eqnarray}
H_1 &=&\int d^2x
\tr\left\{|D_0\phi|^2 %^\dagger(D_0\phi)
 %^\dagger(D_i\phi)
+\omega^2\left(\bar\phi P\phi-(\bar\phi P\phi)^2\right)
\right\},\nonumber\\
&=&%\left\{\begin{array}{l} \left[
\int d^2x \tr\left((1-\phi\bar\phi)|\partial_0\phi \mp i \omega P
\phi  |^2 \right)
%+|A_0-i\\bar\phi\partial_0\phi|^2
\pm\omega  Q
 \geq  \ \pm\omega  Q \,,
%\left[\int d^2x
%\tr\left((1-\phi\bar\phi)A_-A_-^\dagger+|A_0-i
%\bar\phi\partial_0\phi|^2
%+2 |D_{\bar z}\phi|^2\right)+\alpha \tr Q\right] -2\pi
%T\end{array}\right.,
\label{inequality1}
\end{eqnarray}
%with $d_\mp \phi \equiv $.
and the remaining part by
\begin{eqnarray}
H_2 =\int d^2x
\tr %^\dagger(D_0\phi)
|D_i\phi|^2 %^\dagger(D_i\phi)
=%\left\{\begin{array}{l} \left[
2 \int d^2x |D_\pm \phi|^2
\pm
2\pi T \geq  \
\pm
2\pi T \,,
%\left[\int d^2x
%\tr\left((1-\phi\bar\phi)A_-A_-^\dagger+|A_0-i
%\bar\phi\partial_0\phi|^2
%+2 |D_{\bar z}\phi|^2\right)+\alpha \tr Q\right] -2\pi
%T\end{array}\right.,
\label{inequality2}
\end{eqnarray}
%One has then
The inequality,
$H \geq \pm 2\pi T\pm\omega Q$, then holds  for any
independent combination of  the first and the second signatures
in front of $T$ and $Q$.
Thus we conclude that
the Hamiltonian is bounded from below by
\begin{eqnarray}
H \geq  2\pi |T|+\omega |Q|\,.
\label{energy}
\end{eqnarray}

%The minimum value of energy per unit space volume is attained for
%the functions satisfying
The saturation of the bound occurs if
the following equations
\begin{eqnarray}
%A_0=i\bar\phi\partial_0\phi,~
\partial_0\phi =\pm i\omega P
\phi,~~D_\pm \phi=0,\label{bps2}
\end{eqnarray}
for any combination of signatures. Namely there are four branches
of BPS equations saturating the bounds.  From this purely
bosonic consideration, the four branches are equally well served
as a set of BPS equations, whose solutions are bounded from below
by their topological and electric charges.

%where the upper/lower signs are respectively for the BPS/anti-BPS
%cases.
%and the anti-BPS equations
%\begin{eqnarray}
%A_0=i\bar\phi\partial_0\phi,~\partial_0\phi+ i\alpha P
%\phi=0,~D_z\phi=0.\label{bps1}
%\end{eqnarray}
%The particular combination of the signatures in the BPS equations
%are to ensure the real saturation of the bound.
%One can check that in the BPS/anti-BPS cases
For instance, with $\partial_0\phi\mp i\omega P\phi=0$,
the Noether charge satisfies
\begin{eqnarray}
\pm Q=  2\omega\int d^2x\tr[(1-\phi\bar\phi)\,
P\phi(P\phi)^\dagger]\geq 0,
\end{eqnarray}
and, in the right side of the inequality of (\ref{inequality1}),
%likewise, in the BPS case with $\partial_0\phi-i\alpha
%P\phi=0$,
%\begin{eqnarray}
% \tr Q\geq 0.
% \end{eqnarray}
%Clearly, the lower bound of the energy is given by
the contribution of the charge $Q$ is always
positive definite upon the saturation of the bound.
Likewise, one can also show that, if $D_\pm\phi =0$,
$\pm T \ge 0$ respectively and, thus, the right hand side of
(\ref{inequality2}) is always positive definite whenever one has
the saturation of the inequality.

Among the four branches in (\ref{bps2}),
only two combinations of signatures, $(-,+)$ and $(+,-)$,
will be shown to be consistent with the remaining
supersymmetries in the next
section. The remaining two  %are non-BPS 
describe nonsupersymmetric solitons
in a strict sense
of remaining %preserved 
supersymmetries.

%%%%%%%%%%%%%%%%%%%%%%%%%%%%%%%%%%%%%%%%%%%%%%%%%%%%%%%%%%%%%%

%%%%%%%%%%%%%%%%%%%%%%%%%%%%%%%%%%%%%%%%%%%%%%%%%%%%%%%%%%%
\section{ Remaining Supersymmetries and Examples}
%%%%%%%%%%%%%%%%%%%%%%%%%%%%%%%%%%%%%%%%%%%%%%%%%%%%%%%%%%%%%

In this section, we would like to show that the
time-dependent
Q-ball of the previous section preserve 1/2 of the
supersymmetries for some particular branches.
Since the fermionic part of solutions are assumed to
vanish, the variation of the fermionic part has to be zero
for any such solutions preserving some
supersymmetries.
Thus  %show that the above time-dependent solutions are
%$\frac{1}{2}$-BPS states. Let us consider
we would like to show that the variation of the fermionic component,
\begin{equation}
\delta\psi=\left(-i\gamma^0D_0\phi+
\omega (P\phi-\phi\bar\phi
P\phi)\right)\epsilon-i\left(\gamma^1D_1\phi+\gamma_2D_2\phi\right)
\epsilon\,
\label{bpsequation}
\end{equation}
vanishes for some nontrivial constant $\epsilon$. The remaining
supersymmetries are parameterized by
$\epsilon_{\pm}=p_{\pm}\epsilon$ where we introduce the projection
operators by
\begin{equation}
p_{\pm}=\frac{1\mp i\gamma^1\gamma^2}{2}
%p_{\pm}^2=p_{\pm},~-\sigma_2p_{\pm}=\pm p_{\pm},
\label{proj1}
\end{equation}
that satisfies $p_{\pm}^2=p_{\pm}$.
%we define
%\begin{equation}
%p_{\pm}\epsilon=\epsilon_{\pm}.\label{proj2}
%\end{equation}
Then, for this remaining space, the fermionic variation becomes
\begin{equation}
\delta\psi=\left(i(1-\phi\bar\phi) (\pm \partial_0\phi +i \omega
P\phi) -i\gamma^1(D_1\phi\mp i D_2\phi) \right) \epsilon_\pm\,.
\label{bpsequation1}
\end{equation}
Thus, for this expression to be zero  for nontrivial $\epsilon_+$/
$\epsilon_-$,
one needs
\begin{eqnarray}
%A_0=i\bar\phi\partial_0\phi,~
\partial_0\phi =\mp i\omega P
\phi,~~D_\pm \phi=0,
%\label{bps2}
\end{eqnarray}
respectively for the upper/lower combination of signs.
 %the two set of BPS/anti-BPS equations in (\ref{bps2})
These are  precisely the two  branches of $(+,-)$ and $(-,+)$
combinations in (\ref{bps2}). Thus we verified that only the two
branches are really consistent with the remaining supersymmetries.

%%The second term of Eq. (\ref{bpsequation}) becomes
%\begin{equation}
%-i\gamma^1\left(D_1\phi-i\gamma^1\gamma^2D_2\phi\right)\epsilon_{\pm}=
%-i\gamma^1\left(D_1\phi\mp iD_2\phi\right)\epsilon_{\pm}.
%\end{equation}
%For $\epsilon_+$ remaining, we have $(D_1-iD_2)\phi=D_z\phi=0$,
%and for $\epsilon_-$ remaining, we have $(D_1+iD_2)\phi=D_{\bar
%z}\phi=0$. Also, the first term of Eq. (\ref{bpsequation}) is
%given by
%\begin{equation}
%\left(-i\sigma_2D_0\phi+\alpha (P\phi-\phi\bar\phi P\phi)
%\right)\epsilon_{\pm}=\left(\pm iD_0\phi+\alpha
%(P\phi-\phi\bar\phi P\phi)
%\right)\epsilon_{\pm}=(1-\phi\bar\phi)(\pm i\partial_0\phi +
%i\alpha P\phi)\epsilon_{\pm}.
%\end{equation}
%Therefore, we have BPS equation (\ref{bps2}) for non-vanishing
%$\epsilon_-$ and anti-BPS equation (\ref{bps1}) for non-vanishing
%$\epsilon_+$.

The solutions involve the nonzero electric charges together
with the magnetic vortex charge. Due to the BPS equation, the
solution has to be time dependent. The time dependence can be solved
rather trivially by
\begin{eqnarray}
P\phi= e^{\mp i \omega t} P\phi_0 (x_1, x_2)\,,\ \ P_\perp\phi
=P_\perp\phi_0(x_1, x_2)
\end{eqnarray}
where $P_\perp= 1-P$ and $\phi_0$ is time independent.
 Thus we have shown that the supersymmetry
can be preserved even for the configurations with an explicit time
dependence. The solitons here are the  supersymmetric Q-balls, whose
description involve an explicit time dependence inevitably.

 Note that the BPS solutions have nonvanishing electric
and magnetic fluxes.  It is worthwhile to give an explicit example,
and we concentrate on the simple case of $CP(n)$, which corresponds
to the Grassmannian manifold of $ M=1$ and $N=n$. Then explicitly
one has $\phi^T=(\phi_1,~\phi_2,~\cdots, \phi_n, \phi_{n+1})$ and
chooses the projection matrix $P$  by
\begin{eqnarray}
 P={\rm diag}(1,~1,~1,\cdots,1,~0)
 \end{eqnarray}
Let us introduce the projective coordinate $\xi$ via
\begin{eqnarray}
\phi^T=\frac{1}{{\sqrt{1+\vert\xi\vert^2}}}\left(
 \xi_1,
  \xi_2, \cdots, \xi_{n}, 1 \right)
\end{eqnarray}
%We find that the solutions are given by
The first equation of (\ref{bps2}) is solved by
\begin{equation}
%\begin{array}{ll}
\xi=\exp^{\pm i\omega t}\xi_0\,, %&\mbox{:\,BPS~equations}\,,\\
%{}&{}\\
%\xi=\exp^{i\alpha t}\xi_0(z)&\mbox{:\,anti-BPS~equations}\,.
%\end{array}
\label{BPSeq2}
\end{equation}
where $\xi_0$ is time independent. The second equation is
solved then simply by demanding $\xi_0$ as antiholomorphic/holomorphic
function ( i.e. $\xi_0(\bar{z})$/$\, \xi_0({z})$ )
for $+/-$ signatures respectively.
The corresponding  topological charge $T$ and the Noether charge $Q$ are
expressed by
\begin{eqnarray}
T &=& \frac{i}{2\pi}\displaystyle{\int}{\rm d}^{2}x\,
\frac{\partial_{x_1} \xi^\dagger \partial_{x_2} \xi- \partial_{x_2}
\xi^\dagger\partial_{x_1}
\xi}{(1+\xi^\dagger\xi)^2},\\
Q &=& i\displaystyle{\int}{\rm d}^{2}x\, \frac{\partial_t
\xi^\dagger  \xi- \xi^\dagger\partial_t
\xi}{(1+\xi^\dagger\xi)^2}. \label{atr3}
\end{eqnarray}
We note that  the above solution reproduces
 the previous Q-lump solution
of Refs. \cite{lees, abra}.

\section{ Time Dependent Q-balls with Multi Frequency}

The potential in (\ref{potential}) can be extended further by
considering
\begin{eqnarray}
W(\Phi)=%\int  d^2 \theta
\frac{1}{2}\tr~\bar\Phi
M\Phi,\label{potential1}
\end{eqnarray}
where $M=\omega_k P_k$ with $P_kP_k=P_k$ and
$P_kP_l=P_k \delta_{kl}$.
 Then after eliminating the auxiliary fields, the
action with ${\cal N}=2$ supersymmetries
 becomes
\begin{eqnarray}
%V(\phi)=\bar\phi M^2\phi-(\bar\phi M\phi)^2.
%the actions becomes
%Substitution into $S_t$ yields
S &=&\int d^3 x %\frac{N}{g}
\tr\left\{\vert D_\mu\phi\vert^2 +i\bar\psi\gamma^\mu D_\mu\psi
+\left(\bar\phi M\phi-\frac{1}{2}\bar\psi\psi\right)^2-
\bar\phi M^2 \phi+ \bar\psi M\psi\right\}\nonumber\\
&=&\int d^3 x %\frac{N}{g}
\tr\left\{ \vert\partial_\mu\phi\vert^2 +i\bar\psi\gamma^\mu
\partial_\mu\psi
-\left(i\bar\phi\partial_\mu\phi+\frac{1}{2}\bar\psi\gamma_\mu\psi\right)^2
+\left( \bar\phi M\phi-\frac{1}{2}\bar\psi\psi\right)^2-\bar\phi
M^2 \phi+\bar\psi M\psi\right\}.
\end{eqnarray}
Then the analysis of the previous section can be repeated with these
multiple projection operators. Since the analysis is pretty much the
same, we shall present only the result here. One may show that the
Hamiltonian is bounded from below by
 \begin{eqnarray}
H \geq  2\pi |T|+ \sum_k |\omega_k||Q_k|\,.
\label{energym}
\end{eqnarray}
where the charge $Q_k$ is defined by
\begin{eqnarray}
Q_k=-i\int d^2x \tr \left((D_0\phi)\bar\phi P_k
-P_k\phi(D_0\phi)^\dagger\right)\,.
\end{eqnarray}
The saturation of the bound leads to the BPS
equations,
\begin{eqnarray}
%A_0=i\bar\phi\partial_0\phi,~
\partial_0\phi =i \sum_k \epsilon_k  \omega_k P_k
\phi,~~D_\pm \phi=0,\label{bps2m}
\end{eqnarray}
where $\epsilon_k$ can be either $+1$ or $-1$ denoting
independent signatures.

One may further verify that, among these,
only two combinations,
\begin{eqnarray}
%A_0=i\bar\phi\partial_0\phi,~
\partial_0\phi =\mp i \sum_k \omega_k P_k
\phi,~~D_\pm \phi=0,\label{bps2mm}
\end{eqnarray}
are consistent with the remaining supersymmetries.

As before, the time dependence of the BPS solution can be
solved generically; for each sector of  $P_k$, one has
\begin{eqnarray}
P_k\phi= e^{i\epsilon_k \omega_k t} P_k \phi_0 (x_1, x_2)\,,% \
%\bar{P}\phi =\bar{P}\phi_0(x_1, x_2)
\end{eqnarray}
%for each  $k$,
and, for the remaining part,
\begin{eqnarray}
P_\perp\phi =P_\perp\phi_0(x_1, x_2)\,,
\end{eqnarray}
where $P_\perp$ denote now $1-\sum_k P_k$.
% and $\phi_0$ is time independent.
These time dependence is highly nontrivial in the sense that
the sectors of fields defined by the projection $P_k$ interact
with each other nontrivially.

%Then, each $P_i$ sector satisfies the BPS equations %(\ref{bps1})
%(\ref{bps2}) and the energy saturation expression
%(\ref{energy}) is replaced with
%\begin{eqnarray}
%E\geq 2\pi|T|+\Sigma_i \alpha_i|\tr Q_i|.\label{energy1}
%\end{eqnarray}
%for each charge $Q_i$ corresponding to the $P_i$ sector.
%%%%%%%%%%%%%%%%%%%%%%%%%%%%%%%%%%%%%%%%%%%%%%%%%%%%%%%%%%%%%%%%%%%%%%%%%%%%%%%
%%%%%%%%%%%%%%%%%%%%%%%%%%%%%%%%%%%%%%%%%%%%%%%%%%%%%%%%%%%%%%%%%%%%%%%%%%%%
%\section{Extended $Gr(N,m)$ model}
%%%%%%%%%%%%%%%%%%%%%%%%%%%%%%%%%%%%%%%%%%%%%%%%%%%%%%%%%%%%%%%%%%%%%%%%%%%%%%
%%%%%%%%%%%%%%%%%%%%%%%%%%%%%%%%%%%%%%%%%%%%%%%%%%%%%%%%%%%%%%%%%%%%%%%%%%%%

\section{Discussions}
In this note, we construct the ${\cal N}=2$ supersymmetric
Grassmannian nonlinear sigma model for the massless case first and
extend it to the massive ${\cal N}=2$ by adding appropriate
superpotential.
% consistentwith  the  ${\cal N}=2$ supersymmetries.
This massive model allows Q-lump solutions that carry both the
topological and the Noether
charges. We study the corresponding BPS equations %describing these
%lumps
via two methods. One is using the method of  completing
squares in the Hamiltonian, by which one may find that
the Hamiltonian is bounded from below by the charges.
The other is directly finding required conditions for the
solutions to preserve a fraction of supersymmetries using the
supersymmetry variation  of the fermionic part.
These two methods have been giving equivalent sets of BPS equations,
within the present authors' knowledge, for any supersymmetric
theories.  In the present case, however, we find at the end of Section III 
that the former leads to the more general sets of BPS equations
than those from the latter. In a strict sense, the latter
is the precise way of getting supersymmetric solitons,
whose supersymmetric  multiplet has to be short by definition.

These Q-lumps are always time dependent as dictated by the BPS
equations. Thus it is clear that the explicit time dependence can be
consistent with the remaining supersymmetries.
Supertubes\cite{townsend}, for instance, also carry the kinetic
components corresponding to the nonvanishing electric fields but the
solutions are time independent unlike the Q-lump solution discussed
here. We also discuss the supersymmetric Q-lump solution that
involves the time dependence of even multiple frequencies. The
existence of such solutions is highly nontrivial since the sectors
of different frequencies are interacting with each other in a
nontrivial manner.

One curious question is whether the system allows 1/4 BPS equations
or not where the remaining supersymmetry is just one Majorana
component. In case of ${\cal N}=4$ super Yang-Mills theories, the BPS
equations corresponding one Majorana component, i.e. 1/16
supersymmetry, has been constructed\cite{park}.
For some ${\cal N}=2$ theories, 1/8 BPS states are classified
as well\cite{eto}.
%This requies a further study.
Finally, in this note, we did not investigate the detailed
properties
of the supersymmetric Q lump solutions including
their moduli dynamics, interactions and so on. These require  a
further study.

  \acknowledgments
DB would like to thank the Particle Theory Group of 
University of Washington for the hospitality during 
the completion of this work.
  PO like to thank M. Nitta for useful discussions.
  The work of DB was supported in part by
 KOSEF ABRL R14-2003-012-01002-0, and KOSEF SRC CQUeST R11-2005-021.
PO was supported by 
%the Science Research Center Program of the Korea
Science and Engineering Foundation 
KOSEF
through the Center for Quantum
Spacetime (CQUeST) of Sogang University with grant number R11 - 2005
- 021 and by KRF %Korea Research Foundation 
Grant (R05-2004-000-10682-0).

\end{document}